\documentstyle[12pt,aasms4,epsf]{article}
\received{00  1999}
\accepted{00 1999}
\centerline{Submitted to the Editor of the ApJ
 on March 13, 2001}

\def\percc  {\,\mathrm{cm^{-3}} }
\def\iunits {\u erg \ut cm -2 \ut s -1 \ut sr -1 }
\def\ee #1 {\times 10^{#1}}          
\def\ut #1 #2 { \, \textrm{#1}^{#2}} 
\def\u #1 { \, \textrm{#1}}          

\def\deg    {$^{\circ}$}                        
\def\kms    {\hbox{km{\hskip0.1em}s$^{-1}$}}    
\def\msol   {\hbox{$M_\odot$}}                  
\def\dasec  {\hbox{$.\!\!^{\prime\prime}$}}     
\def\asec   {$^{\prime\prime}$}                 
\def\dasec  {\hbox{$.\!\!^{\prime\prime}$}}     
\def\dsec   {\hbox{$.\!\!^{\rm s}$}}            
\def\min    {$^{\rm m}$}                        
\def\hour   {$^{\rm h}$}                        
\def\lsol{\, \hbox{$\hbox{L}_\odot$}}

\def\etal   {{\it et al. }}                     

\begin{document}
\title{High Spectral and Spatial Resolution Observations of 
Shocked Molecular Hydrogen at the Galactic Center}

\author{F. Yusef-Zadeh}
\affil{Department of Physics and Astronomy, Northwestern University,
Evanston, Il. 60208 (zadeh@northwestern.edu)}

\author{S. R. Stolovy}
\affil{SIRTF Science Center, Caltech, Pasadena, CA 91125
(stolovy@ipac.caltech.edu)}

\author{M. Burton}
\affil{School of Physics, University
of New South Wales, Sydney, NSW 2052, Australia 
\& School of Cosmic
Physics, Dublin Institute for Advanced Studies, 5 Merrion Square, Dublin 2, 
Ireland
(M.Burton@unsw.edu.au)}

\author{M. Wardle}
\affil{School of Physics, University of Sydney, NSW 2006, Australia 
(wardle@physics.usyd.edu.au)}

\author{M.C.B.  Ashley}
\affil{School of Physics, University
of New South Wales, Sydney, NSW 2052, Australia (M.Ashley@unsw.edu.au)}

\begin{abstract}

The presence of OH (1720 MHz) masers, and the absence of counterparts
at 1665/1667 MHz has proved to be a clear diagnostic of shocked
molecular gas associated with Galactic supernova remnants.  This
suggests that shocked molecular gas should be associated with the OH
(1720 MHz) masers that have been detected in the circumnuclear disk
(CND) and Sgr A East at the Galactic center.  In order to test this
hypothesis, we observed the H$_2$ 1--0 S(1) and Br
$\gamma$ lines using NICMOS on the HST and UNSWIRF on  the  AAT, near the
regions where OH (1720 MHz) masers are detected in the CND and Sgr A
East.  We present the distribution of H$_2$ in the North  and South lobes of
the CND and in Sgr A East.  
H$_2$ emission accompanies almost all of
the maser spots detected at the Galactic center.  In particular, we
find a striking filamentary structure near the Northwest of the CND and evidence
that shocked molecular gas is associated with the 70 \kms\ molecular
cloud at the Galactic center.  
We argue that the  emission from the CND could arise in gas heated by the
dissipation
of the  random motion of clumps by collisions or
the dissipation of turbulence in a more homogeneous medium. 
In addition,  highly
red-shifted gas of up to 140 \kms\
close to  the eastern edge of the Sgr A East shell is detected.
These observations combined with OH (1720 MHz) results suggest that
the H$_2$ gas is shocked and accelerated by the expansion of Sgr A
East into the 50 and the 70  \kms\ cloud and into the lobes of the  CND.

\end{abstract}


\vfill\eject

\section{Introduction}

The Galactic center provides a unique opportunity to study in detail
the dynamics and physical conditions of the closest galactic nucleus.
The picture that has emerged from multi-wavelength studies of the
Galactic center over the last quarter of a century is that it contains
a clumpy molecular ring (also known as the Circumnuclear Disk, or CND), 
as seen in HCN emission, 
on a scale of 2 to 5 pcs circling the Galactic center with a rotational 
velocity of about 100 \kms\ (see Jackson \etal 1993; Latvakoski \etal  1999 and
references therein).  The CND is heated by UV radiation from the
hot stars  within the central cavity.
The peaks of molecular emission
from the CND are located in its NE (N lobe) and SW (S lobe) and are
consistent with limb-brightening of the inner edges along the
principal axis of a ring.  Within the ring's central cavity, three
``arms'' of ionized gas (Sgr A West) are in orbital motion around the
center (e.g.  Roberts and Goss 1993; Serabyn and Lacy 1985), which is
believed to contain a
$\sim$ 2.5$\times$10$^{6}$ \msol\ black hole (Eckart and Genzel 1996;
Ghez \etal  1998).  The coincidence in the geometry and kinematics of
the southwestern edge of the molecular ring and ionized gas suggests
that the ionized gas is dynamically coupled to the inner edge of the
circumnuclear ring (e.g.  G\"usten et al. 1987).

On a larger scale a non-thermal structure, Sgr A East, is thought to
be a shell-type explosive event, possibly a supernova remnant (SNR),
surrounded by the 50 \kms\ molecular cloud.  A number of observations
present strong evidence that these two objects are physically interacting
with each other (e.g.  Mezger \etal  1996; Zylka \etal  1990;
Serabyn \etal  1992; Yusef-Zadeh \etal  1996).  Thus, the dynamical 
coupling of 
the thermal Sgr A West and the CND, as well as the nonthermal Sgr A East and the
50
\kms\ cloud, has been fairly well established.  A question that will
be addressed  in this paper is the nature of the interaction between these
two systems.

Observations of molecular hydrogen gas are useful, not only for
determining the dynamics of gas, but also for their potential
to distinguish shocked from UV--excited gas.
Molecular hydrogen line emission has been detected from the inner edge
of the circumnuclear ring.  The excitation is thought to be 
produced by
shocks
driven into the ring by the ram pressure associated with the outflow
from the IRS 16 cluster (Gatley \etal 1986; Yusef-Zadeh and Wardle 1993).
Broad emission lines observed toward the Galactic center 
could have contribution   
from 
outflows  from the vicinity of the IRS 16 cluster with a terminal velocity
$v_{w}=500-700$ km s$^{-1}$ and a mass-loss rate $\dot M_{w}\approx
4\times 10^{-3}\;\,M_\odot$ yr$^{-1}$ (Hall, Kleinmann, \& Scoville
1982; Geballe {\it \etal} 1991; Allen {\it \etal} 1990).
However, the detection of shocked (as opposed to UV--heated) gas
associated with the Galactic center molecular ring has been ambiguous.
The intensity ratios of the v=2--1 and 1--0 S(1) lines of molecular
hydrogen (Gatley \etal\ 1986; Burton \& Allen 1992, 1993; Ramsay-Howatt \etal
1993; Pak \etal  1996) are taken to be consistent with collisional
excitation rather than fluorescence from low-density gas, but the high
density of the molecular gas at the Galactic center and the intense UV
radiation field in the region allow the line ratios from UV-irradiated
gas to resemble those of shock-heated gas (Sternberg \& Dalgarno 1989;
Burton \etal 1990).

The OH molecule has recently proved to be useful in
distinguishing shocked from radiatively excited molecular gas through
the presence of 1720 MHz masers.  The presence of the OH (1720 MHz)
maser line, and the absence of the 1665/1667 MHz lines provides a
clear diagnostic of shocked molecular gas, as the far--IR radiation field
from warm dust in UV-heated clouds would pump the latter transitions
(Frail, Goss \& Slysh 1994; Lockett, Gauthier \& Elitzur 1999).
Recent detection of diffuse X-ray emission from 
the interior of Sgr A East and from 
Sgr A West are  consistent with the shock model enhancing the 
abundance of OH behind the shock front (e.g. Baganoff \etal 2001; 
Maeda \etal 2001; Wardle
1999).  The expansion of a supernova remnant into the Sgr A East molecular
cloud (i.e.  the 50 \kms\ cloud) is believed to be responsible for the
production of the OH (1720 MHz) maser emission (Yusef-Zadeh \etal
1996, 1999a).  If the OH (1720 MHz) masers signify regions of shocked
gas, this suggests that the H$_2$ emission is also shock-excited.

In order to investigate this hypothesis, observations of the H$_2$
1--0 S(1) and Br $\gamma$ lines were carried out using the University
of New South Wales Infra-Red Fabry-Perot (UNSWIRF) on the AAT and
NICMOS on the HST near the regions where OH (1720 MHz) masers are
detected in the CND and Sgr A East.  
The NICMOS data provide subarcsecond
spatial resolution and the UNSWIRF data provide velocity
information for both the molecular and ionized gas. 
Preliminary results of these
observations combined with low-frequency radio continuum images 
were
presented by Yusef-Zadeh \etal (1999b) 
who argued for the interaction of Sgr A East and the CND.  

This paper presents the distribution of H$_2$ gas in the CND in section 3.1
and and discusses the nature of H$_2$ gas in section 3.2 followed by the
correlation of H$_2$ linear feature, OH (1720 MHz) masers, the Sgr A
East molecular cloud and radio continuum emission from 
Sgr A East  with each other in
section 3.3 and 3.4. In particular,
the evidence is presented that the H$_2$ linear filament to the NW of the
CND is not only excited by Sgr A East but also is associated with the CND
by a ridge of molecular gas.  We then present the evidence that all
Galactic center OH (1720 MHz) masers, with one exception, are accompanied
by shocked H$_2$ gas with the implication that Sgr A East is responsible
for shocking the gas in the CND, the 50 and the 70 \kms\ molecular clouds. 
In section 3.5, two extinction clouds are discussed toward the Northern Arm
and Sgr A East before conclusions are drawn in section 4.

\section{Observations}

Camera 3 of NICMOS on the HST was used to observe the 1-0 S(1)
transition of H$_{2}$ toward Sgr A West as well as Sgr A East.
Twelve adjacent pointings, each with a 52'' field of view and 0.203''
pixel scale, were observed in 1\% bandwidth filters containing
the line+continuum (F212N) and continuum (F215N). 
The observations were made on July 3 and 4, 1998 
when the plate scale for camera 3 was
0.20384'' x  0.20310'' along the detector axes in x and y,
respectively. Because the majority of the images had to be rotated
by $\sim$ 135 degrees to orient north up, the x and y plate
scales were effectively equally interpolated to an average value of
0.2035''/pixel without the need to correct for non-square pixels.
The individual
pointings were additionally dithered 4 times in a square pattern spaced by
16$''$ along the detector axes in order to provide spatial overlap, correct
for bad pixels, and to improve the sampling. Final mosaics covering
$\sim 4'\times4'$ were constructed by combining all 48 frames
in each filter. Each of the 48 positions was observed for 128 seconds
using a MULTI-ACCUM sequence to correct for cosmic rays and non-linearity.
The standard STScI procedure "calnica" version 3.2 was used to reduce the data,
which does a bias subtraction, linearity correction,
dark subtraction and flat field.
A guide-star reaquisition during the 6-orbit program forced the spacecraft
orientation angle to change during the observation and 8 of the 48 frames in
each filter had to be rotated
with respect to the rest.  An IDL routine developed by the NICMOS team at the
University of Arizona was used to make the mosaic in each filter,
and overlapping good pixels were median combined. Bicubic interpolation
was used to shift and rotate the images. The effective 
plate scale is 0.2035$''$. 

The NICMOS H$_{2}$ image is constructed,  in principle, by subtracting an
appropriately flux-calibrated F215N mosaic from the F212N mosaic.
However, a simple subtraction did not yield a satisfactory line image and
additional steps were taken, as described here. The thermal background
(as measured in regions devoid of stars due to high extinction) was removed from
the data;
DC levels of 0.61 adu/s and 0.71 adu/s were subtracted for
F212N and F215N, respectively. Random fluctuations in DC levels by quadrant in
the NICMOS detectors contributed to another (much fainter) DC background
that needed to be subtracted in rectangular patches; this subtle effect only
was apparent in the F212N-F215N difference image and could not be perfectly
corrected due to the contamination by the stars that dominate both mosaics.
Next, the F215N images were scaled to the F212N image in
order to minimize the stellar residuals upon subtraction; the scale factor used
was 1.02. Both positive and negative stellar residuals remained, however, 
due to intrinsic variations in stellar colors (mostly due to patchy extinction)
and due to PSF differences in alignment and illumination of the undersampled
pixels.  In order to remove the high-spatial frequency residuals and
to improve the signal/noise, a 5x5 pixel median filter was applied   
to the difference image, excluding very negative pixels from the median   
procedure.  The resultant spatial resolution obtained after medianing
was estimated by measuring the effect of the same median on the continuum
image: stars were measured to increase in FWHM from 0.5'' to 0.7''.
The reduction in flux of individual bright pixels in the H$_{2}$ image
due to the median filter was measured to be not more than a factor of 2.5. The
flux
in 1'' apertures remained constant after applying the median filter.
Finally, the difference image was converted to line flux units   
with the calibration supplied by STScI. 
Due to the leakeage of blueshifted Br gamma
emission into the 2.15 $\mu$m continuum filter, portions of
the ``mini-spiral" of ionized gas within the inner parsec show up as
negative in the H$_{2}$ image (masked to zero is equivalent to white in the
H$_{2}$ image).
The final NICMOS H$_{2}$ mosaic is shown in reverse grayscale in
Figures 2a-b, 3 and 7 and  the features
discussed in this paper are labelled in Figure 2b.
Registration of the NICMOS data with the radio data was achieved by
aligning position of IRS 7, the brightest continuum source in
the NICMOS image, with its radio maser position.  The position
of Sgr A* is taken to be that given in Menten \etal  1997. 
Alignment accuracy is estimated to be
correct to better than 1 NICMOS  pixel (0.2$''$).

Observations of the 2.122 $\mu$m line of H$_2$ were also obtained with
the Anglo-Australian Telescope (AAT) in June 1998, using the IRIS
1--2.5$\mu$m camera in conjunction with the
UNSWIRF\footnote{University of New South Wales InfraRed Fabry-Perot}
Fabry-Perot etalon (Ryder \etal 1998).  With a FWHM spectral
resolution of $\sim 75$ \kms, a pixel size of $0.77''$ and a $100''$
circular field of view, the etalon is scanned through a spectral line
of interest with a 40 \kms\ plate spacing, and with an `off-line'
setting chosen to provide continuum subtraction.  Sky frames are also
taken for each etalon spacing.

Four positions were observed, F1 to F4, with central positions
(17\hour\ 42\min, -28\deg\ (1950)) of (30\dsec, 58' 45''), (28\dsec,
59' 40''), (33\dsec, -29\deg\ 00' 10'') \& (34.5\dsec, 59' 50'')
respectively. These are shown in Figure 1, overlaid on a K--band
(2.15$\mu$m) continuum image of the Galactic center from NICMOS.
A mosaic of the four frames was carried out using  FLATN in 
AIPS and the grayscale image of integrated line emission 
is  displayed  in Figure 2c.    
 6, 11, 6
\& 7 plate spacings were taken at each
position, covering 200, 395, 200 \& 240 \kms, respectively.
Integration time per frame was 2 minutes.  Seeing varied between 2''
and 3'' during the observations.

Data reduction was through a custom software page using
IRAF\footnote{Image Reduction and Analysis Facility (www.iraf.noao.edu)}.
Frames are
linearised, flat-fielded using a dome flat, sky-subtracted, shifted to
align the stars in each frame, smoothed and the off-line frame
subtracted from each on-line frame (each having been appropriately
scaled to minimize residuals from the subtraction process).  Stacking
the frames yields a data cube, which is then fitted pixel-by-pixel
with the instrumental profile (a Lorentzian) to yield a line image.
Furthermore, the line center is also determined from the peak position
of the fitted line.  The data cube obtained is
not a true velocity-position cube as the wavelength associated with
each pixel (as well as the spectral resolution) varies across the
array (by up to 40 km/s), but the variation can be calibrated by
observation of an arc line, and hence the line center at each pixel
determined.  Typically this is accurate to $\sim 10$ \kms, but in the
crowded environment of the Galactic center continuum subtraction was
imperfect and  accuracy varies across the field.  Absolute
wavelength calibration is then made by comparison to the etalon
setting for the H$_2$ emission in known sources, in this case IC 4406
(at $-41$ \kms\ V$_{LSR}$) and M17 (+21 \kms\ V$_{LSR}$). The
overall accuracy in the line center determination is estimated 
to be
$\sim 25$ \kms.
The moderate
spectral resolution, and the wide wings of the instrumental profile,
preclude any attempt to determine further spectral information
other than the line center. Flux calibration was made by comparison to the
star HR 6378 with a K--band magnitude of 2.29 magnitudes, and the
absolute accuracy is typically around 30\% in the line fluxes.
An image was also obtained in the 2.166$\mu$m Br $\gamma$ hydrogen
recombination line in June 1999, with a widely-spaced velocity-channel
separation of 95 \kms.  While a detailed discussion of these results will be
given elsewhere, selected images from this data are used in this paper.

\section{Results and Discussion}

\subsection{H$_2$ Gas in the Circumnuclear Ring}

The region observed with UNSWIRF is shown in Figure 1, with the four
positions marked by circles overlaid on a 2.2$\mu$m NICMOS continuum image of
the Galactic center.  
Figure 2a shows the
H$_2$ image obtained with NICMOS, overlaid with the outermost contour
from the 88.6\,GHz J=1--0 line map of HCN (G\"usten \etal 1987), which
delineates the extent of dense gas in the CND\@.  The H$_2$ emission is 
shown in reverse grayscale. 
Also marked are the positions of OH (1720 MHz)
masers (from Yusef-Zadeh \etal 1996) and Sgr A$^*$. 
Although the median filtering which was applied twice removed many of the
stellar
residuals due to the undersampled pixels, some artifacts remain.
Nevertheless, the overall spatial distribution of the H$_2$ emission
is clear, and many individual features are clearly resolved. 
Figures 2b,c  present the prominent H$_2$ features labelled on an NICMOS
H$_2$ image and the mosaic of four UNSWIRF images  of H$_2$ line intensity, 
respectively.  The
strongest H$_2$ emission is associated with the NE and SW lobes of the
CND, as first mapped by Gatley \etal\ (1984, 1986).  The peak H$_2$
intensities in the NE and SW are 7 and
 $8 \times 10^{-15}$  erg s$^{-1}$ cm$^{-2}$ arcsecond$^{-2}$, 
respectively.
Apart from the CND, the
most striking feature is a linear filament to the NW of the CND\@.
Centered on $\alpha = 17^h 42^m 28.5^s, \delta = -28^0
58' 30''$, and extending about $1'$ in a NE--SW direction, the FWHM of
the width of the filament
is only $\sim 1.3''$ (0.05pc at the distance of 8kpc).  
Several compact
knots of emission can be found along the length of the filament.
The most pointlike of these is found at the filament's SW end and
is located 24.8$''$ W, 28.3$''$ N of Sgr A*.  As there is no detected
stellar counterpart in the continuum filter to this compact H$_{2}$ knot,
it is clearly not an artifact of improper stellar subtraction;
therefore it is likely nonstellar in origin despite its pointlike appearance.
The typical
intensity of the H$_2$ emission along the linear  feature is between  
$\sim 2-3 \times 10^{-15}$ 
erg s$^{-1}$ cm$^{-2}$ arcsecond$^{-2}$.
There is a gap in the filament near $\alpha = 17^h 42^m
29^s, \delta$ = -28\deg\ 58' 30''.

Contours of radio continuum emission at 6\,cm obtained with the Very Large
Array of the National Radio Astronomy Observatory\footnote{The National
Radio Astronomy Observatory is a facility of the National Science
Foundation, operated under a cooperative agreement by Associated
Universities, Inc.}  (Yusef-Zadeh \& Morris
1987; Yusef-Zadeh \& Wardle 1993) are superimposed on the NICMOS H$_2$
emission in Figure 3.  The distribution of ionized gas near the N and S
lobes is asymmetric and the spiral-shaped structure of Sgr A West is
evident near the center of the image.  Figure 4 shows the distribution of
molecular hydrogen emission overlaid with contours of Br$\gamma$ emission
extracted from the data cube at 125 \kms\ and $-160$ \kms.  As
expected, the distribution of Br$\gamma$ line emission broadly follows the
radio free-free emission, the lowest contour following the edge of the S
lobe of the CND in Figure 4b but avoiding the N lobe in Figure 4a.  The
intensity of Br$\gamma$ emission is stronger in the N than the S lobe by a
factor of about 3, peaking at $2 \times 10^{-14}$
erg s$^{-1}$ cm$^{-2}$ arcsecond$^{-2}$, 
in the N and $6 \times 10^{-15}$
erg s$^{-1}$ cm$^{-2}$ arcsecond$^{-2}$, 
in the S\@.  
The brightest Br $\gamma$ emission arises from
the mini-cavity with the flux of $3\times10^{-14}$
erg s$^{-1}$ cm$^{-2}$ arcsecond$^{-2}$
(Roberts, Yusef-Zadeh and Goss
1996)  which lies near Sgr
A$^*$.  The emission velocities for the Br$\gamma$ features peak at -350,
+29 and -160 for the mini-cavity, N. arm  and S. arm of Sgr A West,
respectively (A more
detailed discussion of Br$\gamma$ data will be given elsewhere). 


Figure 5a shows the distribution of the integrated intensity of the
H$_2$ 1--0 S(1) line for Field 1 (F1), which includes the northern
part of the CND and the newly discovered filament to its NW\@.  Figure
5b shows the same image superimposed on a 2.2$\mu$m continuum image,
and also has the line center velocities for the principle emission
features labelled, in addition to the OH 1720\,MHz masers. The
brightest clump in the N lobe, with V$_{LSR}$ velocity of 110-115
\kms, has peak intensity of $6.6 \ee -15 \u erg \ut cm -2 \ut s -1 \ut
arcsec -2 $.  The 134 \kms\ OH maser (B) lies adjacent to the western edge
of
the N. lobe. 

The H$_2$ linear filament is brightest at $\approx$ 75 \kms,
but the line center velocity decreases to $\sim 50$ \kms\ along its NE
extension.  The western edge of the CND (sometimes called the western
Arc) appears at velocities ranging between 55 and 80 \kms.  The OH
(1720 MHz) masers in this vicinity are distributed along the edges of
the N lobe and the filament where the intensity of H$_2$ emission
falls from the peak values, to levels of 1--2 $\times 10^{-15} \u erg
\ut cm -2 \ut s -1 \ut arcsec -2 $.  At the position of the 43 \kms\
maser (C) a ridge of diffuse H$_2$ emission appears to connect the
filament to the western edge of the CND\@. However, the interpretation
at this location is confused by the presence of the ``70 \kms\ cloud''
seen in [OI] 63$\mu$m (Jackson \etal 1993), a feature known to be
associated with the CND\@.  The kinematics of the H$_2$ gas in the
filament and the ridge, with velocities around 60 \kms, suggests that
the H$_2$ emission from these features is indeed associated with the
70 \kms\ cloud.

A plume-like feature within the CND, as shown in Figure 5a, 
also extends
from just northeast
of Sgr A$^{\mathrm{*}}$ near IRS\,7 is located at 
$\alpha = 17^h 42^m 30.2^s, \delta = -28^0 59'3''$. 
  This new feature peaks at
emission velocities near 80 \kms\ with a peak flux of 1.5 $\times
10^{-15} \u erg \ut cm -2 \ut s -1 \ut arcsec -2 $. It does not appear
to be associated with any ionized streamers of Sgr A West but is
adjacent to the [OI] 63\micron\ peak detected by Jackson \etal (1993)
at similar velocities.

H$_2$ emission is also present on the negative Galactic longitude side of the
CND. The maps in Figures 6a and 6b show the line intensity in Field 2
(F2), centered near the southern lobe of the CND\@.  The brightest
clump of the S lobe, with V$_{LSR}$ velocity of $-40$ \kms, has peak
intensity of $7.6 \ee -15 \u erg \ut cm -2 \ut s -1 \ut arcsec -2 $.
Weak, diffuse H$_2$ emission is also present outside and inside the
CND, with intensities $\sim 1 \ee -15 \u erg \ut cm -2 \ut s -1 \ut
arcsec -2 $.  Within the CND, as outlined by the HCN emission (Wright
\etal 1989), there is diffuse H$_2$ emission, at positive velocities.
Diffuse emission also extends beyond the CND, to the SW, with both
negative and positive velocity components.  It is clear that there is
H$_2$ gas within the molecular cavity and that its kinematics on the
negative-longitude side of the CND is inconsistent with the sense of
rotation of the circumnuclear ring.

\subsection{The Nature of H$_2$ Emission from  the Circumnuclear
Ring}

Our UNSWIRF and NICMOS observations detect peak fluxes in the 1-0 S(1)
molecular hydrogen line of approximately $7\ee -15 \u erg \ut s -1 \ut
cm -2 \ut arcsec -2 $ in the N and S lobes.  This is comparable to the
fluxes reported by Gatley \etal (1986) ($\approx 4\ee -15 \u erg \ut s
-1 \ut cm -2 \ut arcsec -2 $ in an 18\asec\ diameter aperture) but
somewhat more than by Burton \& Allen (1992) ($\approx
2\ee -15 \u erg \ut s -1 \ut cm -2 \ut arcsec -2 $ in a 1\dasec 4 by
4\dasec 7 EW aperture placed, but not peaked,  on the N lobe).  Adopting
$A_K =3$,
the extinction-corrected line intensity from the lobes is approximately 
$4\times10^{-3} \u erg \ut s -1 \ut cm -2 \ut sr -1 $ at its peak.
The extinction, of course, could be higher toward the N. and S. lobes
as recently discussed by Stolovy, Scoville \& Yusef-Zadeh (2001). 

The distribution of emission is broadly consistent with the 
orientation of the inner circumnuclear disk inferred from HCN 
observations (G\"usten \etal 1987; Jackson \etal 1993): a 0.5 pc thick 
torus of emission with normal surface intensity $\sim 10^{-3} \iunits$ 
inclined at 70 degrees to the line of sight will produce the N and S 
lobes by limb brightening.  The source of excitation of the H$_2$ 
emission is puzzling.  Gatley \etal (1984) argued that the inner edge 
of the circumnuclear ring is shocked by a wind from the mass-losing 
stars in the central few pc, for which $\dot{M}\sim 3\ee -3 \msol\ut 
yr -1 $ and $v_{\mathrm{wind}}\approx 750 \u km \ut s -1 $ (Krabbe 
\etal 1991; Najarro \etal 1997).  At 1.7 pc, this combined wind would 
be capable of driving a shock of speed $v_s$ into a pre-shock medium 
of H density $n_H$ with $n_H v_s^2 \approx 2\ee 6 \ut cm -3 \ut km 2 
\ut s -2 $.  However, the observed intensity can be produced by C-type 
shock waves with $v_s \approx 30 \kms$ if the pre-shock density is 
$\sim 10^4 \ut cm -3 $, or in both C- and J-type shocks with $v_s 
\approx 20 \kms$ if the pre-shock density is $\ga 10^5 \percc$ (Kwan 
1977; Draine, Roberge \& Dalgarno 1983; Kaufman \& Neufeld 1996a).  
Thus the ram pressure of the IRS 16 wind is at most one fifth of that 
required.\footnote{Gatley \etal (1984) conclude that this ram pressure 
is sufficient, but this is based on an 
estimate that the $\sim 50 \lsol$ emitted
in the 
1--0 S(1) line arises from $\sim 0.01\msol $ of gas at 2000 K, which 
appears to be a factor of ten too low.}

The intense UV field ($G_0\sim 10^5$) at the Galactic center 
(Sternberg \& Dalgarno 1989; Burton, Hollenbach \& Tielens 1990) is 
another source of excitation, although UV may be prevented from 
reaching the eastern side of the ring by intervening material 
associated with the Northern Arm (Genzel, Hollenbach \& Townes 1995).  
In equilibrium, a PDR emits $\sim 10^{-3}\iunits$ in the 1--0 S(1) 
line only if the gas density $\ga 10^7 \percc$, but then the predicted 
[OI] 63\micron\ intensity ($\ga 0.1 \iunits$; Burton \etal 1990), is 
several times higher than observed (Jackson \etal 1993).  Although the 
presence of high-density gas ($n\sim 10^6$--$10^8\ut cm -3 $) at the 
inner edge of the circumnuclear ring has been inferred from HCN 
observations (G\"usten \etal 1987, Jackson \etal 1993, Marshall, 
Lasenby \& Harris 1995), this material has a sky covering fraction 
$\sim 0.1$, and so the H$_2$ line intensity would be diluted by this 
factor.  In fact, the observed fluxes in the [OI] 63\micron, [CII] 
158$\mu$m and [SiII] 35$\mu$m far--IR lines are consistent with a 
density $\sim 10^5$ cm$^{-3}$ and covering fraction $\sim 1$ (Burton 
\etal 1990; Wolfire, Hollenbach \& Tielens 1990), consistent with UV 
heating of the \emph{envelopes} of the dense cores seen in HCN 
(Jackson \etal 1993).  In equilibrium, the UV-heated envelopes would 
produce a flux of only $\sim 10^{-5}\iunits$.  The H$_2$ emission from 
the envelopes is increased substantially if their exposure to the UV 
flux varies on a timescale $\la 300\u yr $(Goldshmidt \& Sternberg 
1995; Hollenbach \& Natta 1995).  In this case, the 1--0 S(1) line 
intensity is $\sim 10^{-3}\iunits$, the 2--1/1--0 S(1) ratio is $\sim 
0.1$, and the far-IR line intensities are still matched because they 
reach equilibrium on a much shorter timescale (Hollenbach \& Natta 
1995).  The only plausible source of variation\footnote{That \emph{we} 
can think of.} on this time scale is the shadowing by other clumps 
that would occur if $f_A \sim 1$ and the clumps have size $\sim 0.01 
\u pc $, assuming that the clump-clump velocity dispersion is $\sim 
30 \kms$.  A similar shadowing effect has been postulated to produce 
intense CI fine-structure emission from dense PDRs in molecular clouds 
(St\"orzer, Stutzki \& Sternberg 1997).

The emission could also arise from gas heated by the dissipation of 
the $\approx$ 30 \kms\ velocity dispersion in the ring (e.g.  Genzel 
1989; Jackson \etal 1993), either by internal shock waves in a 
homogeneous, but turbulent, medium or by collisions in a clumpy 
medium.  The strength of the H$_2$ line emission from clump collisions 
can be estimated as follows.  Characterize the clumps by radius $r$, 
number density $n=n_6\,10^6 \percc$, and clump-to-clump velocity 
dispersion $v = v_{30}\cdot 30 \kms$.  In a collision, assume that a 
shock of area $\pi r^2$ and speed $v$ is driven into each cloud, and 
that a fraction $\varepsilon$ of the mechanical energy flux 
$\frac{1}{2}\rho v^3$ incident on each shock is converted into 
emission in the 1--0 S(1) line of H$_2$.  Using the collision cross
section, 
$\sigma \sim \pi(2r)^2$, and the
duration of a collision, $\sim 2r/v$, then the 
the fraction of clumps that are being shocked at any given time 
is $\approx n_{cl} \sigma v \cdot 2r/v = 8\pi r^3 n_{cl} = 6f_V$, 
where $n_{cl}$ is the number of clumps per unit volume and $f_V$ is 
the volume filling fraction of the clumps.  The column density along 
the $\approx 1 \u pc $ intersection of the line of sight with the  
lobes of the circumnuclear ring inferred from far--IR emission from 
dust grains is $\approx 10^{22} \ut cm -2 $ (Latvakoski \etal 1999).  
This implies $f_V \approx 0.0032 / n_6$, and the line intensity is
\begin{equation}
	I = \frac{\varepsilon\rho v^3}{8 \pi}\, 6f_V f_A \approx 1.0\ee -3
	\varepsilon_{0.02} f_A v_{30}^3 \iunits
	\label{eq:I}
\end{equation}
where $f_A$ is the area covering fraction of the clumps and 
$\varepsilon = 0.02 \varepsilon_{0.02}$.  $\varepsilon$ is an 
initially increasing function of shock speed and density; dropping 
sharply for shock speeds in excess of 40--50 \kms\ when shocks become 
J-type (Hollenbach \& McKee 1989), and decreasing at densities $\ga 
10^8\percc$ because of collisional de-excitation of H$_2$ (Kaufman \& 
Neufeld 1996b) The maximum value, $\sim 0.02$, is obtained for 
shock speeds between $\sim$30 and $\sim 45 \kms $ and $n\sim 
10^5$--$10^{7}\percc$ (Kaufman \& 
Neufeld 1996a).  Thus we conclude that clump collisions are 
capable of generating the observed intensity provided that $f_A\ga 1$ 
and $n\ga 10^5\percc$.

Both the time-dependent UV irradiation and clump collision model 
produce intensities that are roughly density-independent, which may 
explain the uniformity of the H$_2$ emission around the ring.  Both 
scenarios also explain why the large-scale distribution of H$_2$ emission
from the ring traces the 
HCN J=1--0 emission  reasonably well. The lack of obvious small-scale 
fluctuations in the H$_2$ emission implies that several UV-irradiated 
clumps or colliding clumps must be present per square arcsecond.   In 
the colliding clump scenario, this requires that the clump density 
$\ga 10^6 \percc$.

An additional process that could complicate the interpretation 
of  H$_2$ emission from the ring is its physical interaction 
with Sgr A East evidenced by OH (1720 MHz) masers. 
The new class of OH (1720 MHz)  masers which are
also called ``supernova
masers'' are very rare and in all observed 20 sources that have been
reported, the masers are physically associated with supernova remnants (Frail
et al. 1996;  Green \etal 1997; Koralesky \etal 1998;  Yusef-Zadeh \etal
1999c). The most obvious SNR candidate associated with OH (1720 MHz) masers
in the Galactic center is the nonthermal Sgr A East SNR driving a shock
into the CND. Recent analysis of the 1720
MHz maser observations carried out in 1986 shows a -132 \kms\ OH (1720 MHz) 
maser associated with the S. lobe (Yusef-Zadeh \etal  2001; M.  Goss,
private communication). These high-velocity masers give a compelling
evidence that the masers of the CND are produced by the expansion of Sgr A
East and that the H$_2$ molecular emission in the CND is in part shock
excited externally by Sgr A East.  In addition,  the highly blue and
red-shifted OH (1720 MHz) maser features correspond to the systemic
velocity of the molecular gas in the CND. 
 This is because the path of
maximum amplification for inversion of the 1720 MHz OH maser is formed when
the acceleration produced by the shock is transverse to the line of sight
(Frail, Goss and Slysh 1994). This implies that the rotational velocity of
the circumnuclear ring is about 130 \kms.





\subsection{The H$_2$ Linear Filament}


The width of the linear  filament from the NICMOS observations is $\la
1.3''$,
and the H$_2$ 1--0 S(1) line intensity is $\approx 1 \ee -4 \iunits$.  Several
arguments imply that the filament is shock-heated gas rather than from
a PDR\@. Shock models can easily produce the observed intensity and the
H$_2$ emission velocity peaks at 50--75 \kms\ (see Figure 5b), 
which suggests that it is associated with OH (1720
MHz) maser
C at 43 \kms. PDR models, on the other hand, require $G_0 \sim
10^6$, $n_H\ga 10^6 \percc$ and a high inclination to the line of
sight (e.g.  Burton \etal  1990).  The H$_2$ emission in a PDR arises
in a layer of thickness corresponding to $A_V \approx 1$, which would
be $\la 2\ee 15 \u cm $, corresponding to an angular scale $\la
0\dasec 02$.  Further, there is no evidence for the associated
ionization front as seen both in high-frequency radio continuum 
and
Br$\gamma$ line images.
Interestingly, the H$_2$ filament lies along the western  edge of the 
Sgr A East shell (Yusef-Zadeh \etal 1999b) which is known to be 
nonthermal. 
Figure 7 shows the distribution of H$_2$ line emission based on NICMOS
observations superimposed on a contour of 20cm emission from  the Sgr A
East shell. 
The alignment
of the western edge of the nonthermal Sgr A East 
shell, the linear H$_2$ filament and maser C  
is used
as a
compelling evidence that the H$_2$ filament and the OH (1720 MHz) maser C 
are tracing shocked molecular gas produced by the expansion of Sgr A East
into the western edge of the 50 \kms molecular cloud.

\subsection{H$_2$ Emission and OH (1720 MHz) Masers}


Figure 2a shows the large-scale view of H$_2$ emission from
NICMOS observations outlined by contours of HCN emission.  A higher
concentration of stars to the northeast is likely to be due to the Sgr A
East 50 \kms\ cloud absorbing the background stellar continuum.  Figure 2
shows a number of H$_2$ emitting clouds  beyond the contours of HCN
emission.  We describe below individual sources of H$_2$ emission.


The region near 
$\alpha = 17^h 42^m 32^s, \delta = -29^0 00'30''$ 
to
the southeast of Figure 2b is of particular interest because a number
of OH (1720 MHz) masers have been detected there (Yusef-Zadeh \etal 
1996).  The
locations of the OH (1720 MHz) maser features A and D--G (velocities
of +50--60 km/s) within the M-0.02-0.07 (``50 km/s'') cloud are
indicated in this Figure. 
Spectroscopy toward these sources
showed evidence of weak H$_2$ emission lying close to the position of 
these masers
(Wardle, Yusef-Zadeh and Geballe 1999).  Figure 8 (F3 in Figure 1) 
shows the integrated H$_2$ 
1--0 S(1) emission towards these masers.
Similarly, Figure 9 shows the H$_2$ distribution 
in an overlapping pointing 
(F4 in Figure 1).

The masers in Figures 8 and 9 appear to be aligned along an elongated
nonthermal
continuum feature at the southeastern boundary of Sgr A East at 20\,cm
(Yusef-Zadeh and
Morris 1987; Yusef-Zadeh \etal 1996).
The H$_2$ gas which appears to be associated with masers A--E (those 
labelled 66 and 57 \kms) has a
flux density 5--10 $\times 10^{-16} \u erg \ut s -1 \ut cm -2 \ut arcsec -2 $
and a mean velocity of about 95 \kms.  NH$_3$ emission between
$-41$ and 101 \kms\ (Coil and Ho 1999) also lies adjacent to the maser
sources, but is not seen associated with the NS elongated H$_2$ feature
centered at $\alpha = 17^h 42^m 32.7^s, \delta = -28^0
59' 45''$.  
The G OH (1720 MHz) maser (labelled by 55 \kms) does not appear to have have any 
detectable H$_2$ counterpart. 
While the velocities of the OH masers and the corresponding H$_2$ emission
are not the same, differing by $\sim 30$ \kms\ (as is the case for the OH masers
in the northern lobe of the CND), they do support a physical association
between them.
The poor spatial correlation between OH masers and 
NH$_3$ gas is probably
explained by the restricted
physical conditions under
which OH (1720 MHz) masers can be formed (Lockett \etal  1999) as well as the
fact that the velocities of  OH (1720 MHz) masers trace 
 the systemic motion of
the molecular clouds.
A number of high-velocity   H$_2$ features 
are also noted lying within and  at the boundary of the Sgr A East shell
(centered at 
$\alpha = 17^h 42^m 32.7^s, \delta = -28^0 59' 35''$
and 
$\alpha = 17^h 42^m 37^s, \delta = -28^0  59' 50'$.)
These are best shown in Figure 9  and have line
center velocities ranging between 85 and 140 \kms.  

Two prominent  H$_2$ clouds are also noted in 
projection  against the interior of the Sgr A East shell
and to the north of the N. lobe 
near $\alpha = 17^h 42^m 33^s, \delta = -28^0 58' 15''$
These features which  are  noted in Figures 2a,b, 7  are
located about
90$''$ NE of 
Sgr A$^*$ and are labelled as "Outer H$_2$ Clumps" in Figure 2b.. 
The 20cm image of this field shows
a nonthermal ridge of  emission at this location. The morphology of
this ridge with respect to the distribution of the CND was
recently
discussed (see the supplement figure of Yusef-Zadeh, Melia and
Wardle 2000) as possible evidence for the
physical interaction of the CND and Sgr A East.
The velocity structure of these H$_2$ clouds are unknown, 
lying outside the UNSWIRF  fields, thus
their physical association with either Sgr A East or the CND is 
unclear. 

\subsection{The Outer H$_2$  Filament and Clumps}

Another new straight filamentary structure which appears somewhat broader
and weaker than the linear filament is shown in Figures 2a,b. The typical
H$_{2}$ flux uncorrected for extinction ranges between 0.7 and 1.7 $\times
10^{-15}$ erg s$^{-1}$ cm$^{-2}$ arcsecond$^{-2}$, 
This feature is oriented
roughly in the N-S direction and extending well beyond the inner edge of
the CND.  This "Outer Filament" is seen to extend for at least 50$''$ (and
in fact may extend beyond the boundary of the NICMOS image). This straight
but clumpy Outer Filament is of order 2-3'' in FWHM across 
and is coincident with a similarily elongated structure 
observed in the 34.8$\mu$m  [SiII] line (Stolovy 1997). 
An enhancement in
the [SiII]/dust
continuum in this region also suggests an increase in the atomic silicon
abundance due to grain destruction by shocks. In addition, number of H$_2$
clumps are seen throughout this region, two of which located 50$''$ E and
70$''$ N of Sgr A$^*$.  The "Outer H$_2$ Clumps" to the NE of the northern
lobe appear to be located  between   the inner edge of the nonthermal
shell of
Sgr A East and the outer edge of the CND.

\subsection{Extinction Clouds}


The prominent H$_2$
filament as well as
the N lobe and the western edge of the CND are clearly
visible, in black. 
A number of extinction clouds are apparent throughout
the image (evident in white in the figure).  
Areas of high extinction are apparent in the 
continuum image (Fig. 1) as
regions in white, marking a relative absence of stars, and in the H2 line 
image (Fig 2b, also in white) as regions
where the residuals from imperfect continuum subtraction are least, a
result 
of there being less stars there in the
first place.
Because the stellar flux is weak in its surface brightness at both the
2.12$\mu$m and 2.15$\mu$m
images, the areas of high extinction show up flatly near zero in the
H$_{2}$ image, which appears as white patches in the reverse grayscale.
We discuss below the association of two prominent
extinction clouds with Sgr A East and Sgr A West.

\subsubsection{N. Arm Extinction Cloud}

An  extinction feature in the NICMOS H$_2$ image, with an
angular size of about 20$''$, coincides with the gap in the CND near
$\alpha = 17^h 42^m 30^s, \delta = -28^0 58' 50''$
or about 30$''$ N of Sgr A$^*$ (cf.  Figures 2a,b \& 7).
Several factors suggest that this cloud is associated with the
Northern Arm.  The northernmost tip of the N arm in the 6\,cm data,
near IRS\,8, as seen in Figure 3, coincides with the cloud.  The kinematics
of ionized gas
along the N arm show multiple non-circular velocity components at
$\alpha = 17^h 42^m 29.8^s, \delta = -28^0 59' 00''$, where the N arm
bends in an otherwise continuous velocity distribution (Serabyn and
Lacy 1985). 
Stolovy, Scoville, and Yusef-Zadeh (2000) have recently  measured 
the extinction toward  this cloud at subarcsecond resolution
to be at least A$_v\sim$60 magnitudes.  It is likely higher than this
in the darkest parts beyond the sensitivity of the measurement.
 In addition, the northern half of the N arm also shows a
sudden drop by a factor of 2 in its surface brightness at 6\,cm and a
dearth of [FeII] 1.64$\mu$m emission (Yusef-Zadeh \etal 1999a) to the
north of $\alpha = 17^h 42^m 29.7^s, \delta = -28^0 59' 00''$---ie
where the extinction feature is (Yusef-Zadeh and Wardle 1993).
However, the strongest evidence that the extinction cloud lies within
the CND comes from the $\lambda$1.2\,cm VLA continuum image shown in Figure
10.
This shows the N Arm of Sgr A West and a faint semi-circular shell of
ionized gas surrounding the northern tip of the N arm.  Similarly, a
radiograph of this region (see Fig.\ 4 of Yusef-Zadeh \& Morris 1987),
reveals what appears to be a limb-brightened hole in the distribution
of continuum emission at this location.  The morphology in Figure 10
is suggestive of a neutral cloud surrounded by ionized gas
(Yusef-Zadeh, Zhao \& Goss 1994).  The ionized shell suggests that the
cloud is photoionized externally and is associated with the northern
tip of the N arm.  Indeed, this feature coincides with the outer rim
of the extinction cloud seen with NICMOS at 2.12$\mu$m.  Its surface
brightness at 1.2\,cm, $\sim 0.2$\,mJy within a 0.3$''\times0.2''$
beam, is at the level expected for the ionizing radiation field within
the cavity.  This cloud is possibly the single cloud from which the N
arm originated before falling on a circular orbit toward Sgr\,A$^*$.
Earlier kinematic studies of 12.8\micron\ [NeII] line emission by
Serabyn and Lacy (1985) predicted such a cloud would be colliding with
the CND\@. The cloud has also been detected in an HI absorption study
of this region, where it appears as a feature at 130
\kms\ (Plante, Lo \& Crutcher
1995).  The position, strength and direction of HI Zeeman
splitting measurements at +130 \kms\ agree with the Zeeman study of
the 1720 MHz OH masers at the same location (Plante \etal 1995,
Yusef-Zadeh \etal 1996).

\subsubsection{The Extended  Extinction Cloud}

Another prominent and extended extinction feature is located
to the NW of the Linear H$_{2}$ Filament, extending NE-SW from around
$\alpha = 17^h 42^m 27^s, \delta = -28$\deg 58' 15'' or about 70$''$ NW of
Sgr A$^*$ in Figure 2a,b and 7.
The boundaries of this feature are not
well-defined, but it appears to ``wrap around'' the linear filament and
continue southward, connecting to the Northern Arm extinction 
Cloud
described in the previous sub-section.  The  contour
of  20cm emission in Figure 7 
runs parallel to the inner edge of this
extinction feature. 



The eastern side of Sgr A East is known to be interacting with the 50
\kms\ cloud (e.g.  Mezger \etal 1989; Zylka \etal 1990; Serabyn \etal
1992; Yusef-Zadeh \etal 1996; Coil \& Ho 1999).  However, the
interaction of the western half of Sgr A East has not been examined in
detail before.  The extended extinction cloud  discussed here 
may also be associated with the well-known 50 \kms\ cloud,
assuming that the OH (1720 MHz) maser at 43
\kms\ represents the systemic motion of the cloud at its interaction
site with Sgr A East.

A number of weakly emitting radio continuum features,
known as the streamers, have also been recognized 
running perpendicular to
the CND (Yusef-Zadeh
\& Morris 1987; Yusef-Zadeh \& Wardle 1993).  H110$\alpha$ observations also
indicate gas
velocities up to +144 \kms\ near $\alpha = 17^h 42^m 28^s, \delta =
-28^0 58' 15''$ (just to the NW of the H$_2$ filament).  Along with
the kinematics of the ionized gas near the 43 \kms\ OH maser, these
velocities are inconsistent with circular orbital motion about the
Galactic nucleus.  (Yusef-Zadeh, Zhao \& Goss 1994).  This suggests
that these features in the ionized gas are also associated with the
same molecular cloud producing the extinction feature at 2.12\micron.

\section{Conclusions}

In summary, we have presented H$_2$ 1--0 S(1) line observations of the
circumnuclear ring and Sgr A East using NICMOS on the HST and UNSWIRF on
the AAT. These high spatial and spectral resolution images are correlated
with OH (1720 MHz) masers as well as Br $\gamma$ line and radio continuum
observations of the complex region of the Galactic center. In spite of the
difficulty of detecting shocked molecular gas due to the large intrinsic
linewidth of Galactic center molecular clouds, these observations show
strong evidence of hot shocked H$_2$ gas and cool postshock gas as traced
by OH (1720 MHz) masers.  The interaction of two dynamically coupled
systems associated with Sgr A East and the cicumnuclear ring is discussed.
This argument is based on the fact that in all 20 supernova masers that
have been discovered in the Galaxy, OH (1720 MHz) masers are clearly
evident at the boundary of supernova remnants and molecular clouds where
the interaction is taking place. In the Galactic center region, the
expansion of the nonthermal shell of Sgr A East drives a shock into both
the 50 \kms\ cloud and the circumnuclear ring and produces the observed OH
(1720 MHz) masers. Lastly, It was  argued that the molecular H$_2$ emission
from the circumnuclear ring results from gas heated in the dissipation of
the random motion of molecular clumps in the ring.


\acknowledgments {
We thank 
Ylva Schuberth and Tony
Travouillon for their help in reduction of the UNSWIRF data.
We also thank D. Roberts for useful discussions.  This work was
supported by  a NASA grant GO-07844}




\begin{figure}
\caption{ 2.15$\micron$ continuum image of the Galactic center (from
NICMOS observations) overlaid with the four fields (F1 to F4)
observed in H$_2$ using the AAT/UNSWIRF\@. The circles mark the center
of the 100$''$ diameter frames.  The Galactic plane is apparent
running NE--SW across the image.  
Offsets in arcseconds for circular field centers from Sgr A$^*$
are:  F1 (9.0, 33.3), F2 (-17.2   -21.7), 
F3 (48.4, -51.7),  and F4 (68.3, -31.7). 
The grayscale in the image ranges from
-5 $\mu$Jy pixel$^{-1}$ to 500 $\mu$Jy/pixel. The noise level is 1.5
$\mu$Jy pixel$^{-1}$,
and the peak pixel at IRS7 (which is NOT saturated) is at at level
of 0.12 Jy pixel$^{-1}$ .  The flux of IRS7 at 2.15 microns is measured to
be 0.95 Jy, or 7.2 mag.  Stars as faint as 15th magnitude are
measured in uncrowded regions of high extinction.
The white cross
coincides with the position of Sgr A$^*$.  }
\end{figure}

\begin{figure}
\caption{ (a) Continuum-subtracted H$_2$ 1--0 S(1) line 2.12\micron\ 
image of the Galactic center obtained with HST/ NICMOS\@.  Contours
indicate the extent of the HCN J=1--0 line 88.6\,GHz emission from the
inner region of the circumnuclear ring, as mapped by G\"usten \etal
(1987). The star sign coincides with the position of Sgr A$^*$. 
Despite
the high spatial
resolution of the HST image, and the stability of the point spread
function, this image illustrates the extreme difficulty in obtaining a
line image in a crowded region when only low (1\%) spectral resolution
is used. The H$_2$ emission is shown in reverse grayscale. 
The star  sign corresponds 
with the position of Sgr A$^*$. 
(b) Prominent features are labelled on this NICMOS image of 
H$_2$ line emission. 
The reverse grayscale ranges from -0.05 to 1.5 
 $\times 10^{-16} \u erg \ut cm -2 \ut s -1 \ut pixel -1 $
(1 pixel =0.2035$''$), and is chosen to highlight the fainter features.
The noise level is of order 0.02 
 $\times 10^{-16} \u erg \ut cm -2 \ut s -1 \ut pixel -1 $
and the brightest clumps peak at 3.8 in
these units, yielding a peak S/N of 190. 
The plus signs  show the
positions of the  134 and 43 \kms\ OH (1720MHz) maser B and C,
respectively,  
from Yusef-Zadeh \etal
(1996) as well as the -132 \kms\ maser feature in the S. lobe (Yusef-Zadeh 
\etal 2001). 
(c) Mosaic image of four UNSWIRF frames  (F1 to F4) 
showing grayscale and contours  of integrated   H$_2$ S(1) 1--0   line
emission.
The square in the the ring is  due to artifacts of mosicing.  
(In subsequent figures plus
signs mark the position of OH (1720\,MHz) masers.) 
 }
\end{figure}

\begin{figure}
\caption{Contours of 6\,cm radio continuum emission from Sgr A West,
obtained with a resolution of 3.4$'' \times 2.9''$ and set
at (75, 100, 150, 200, 250, 300, 400, 500) $\times$
1\,mJy, superimposed on a greyscale NICMOS image of the H$_2$ 1--0
S(1) line emission from the CND\@. The star  sign corresponds 
with the position of Sgr A$^*$. }
\end{figure}


\begin{figure}
\caption{\emph{Greyscale:} The left and right panels show 
H$_2$ 1--0 S(1)
emission from
the N and S lobes, in reverse grayscle, 
extracted from two frames of the
UNSWIRF data
cubes
(F1 at 125 \kms\ (N. lobe) and F2 at $-85$ \kms\ (S. lobe), for the two
lobes of the circumnueclear ring,  respectively). The location of 
frames F1 and F2 can be seen in Figure 1.  
\emph{Contours:}
Br $\gamma$ emission extracted from the data cube at  125 \kms\ 
superimposed on the left panel whereas the  
$-160$ \kms\ Br $\gamma$ emission is superimposed on the 
right panel. As discussed in
the text, these are not true velocity
channel frames, but do give an indication of the morphology of the
line emission at these velocities.  Contour levels are (1, 2, 3, 4,
5, 7, 9, 22, 25, 20, 30 and 40) times 5$\times10^{-16}$ erg s$^{-1}$
cm$^{-2}$ arcsec$^{-2}$.
Plus signs  mark
the location of compact OH (1720) masers (source B from Yusef-Zadeh
\etal 1996) associated with the N lobe, emitting at 134 \kms, V$_{LSR}$.
The star sign coincides with the position of Sgr A$^*$.}
\end{figure}

\begin{figure}
\caption{(a) Contours of the integrated 2.12\micron\ 
H$_2$ 1--0 S(1) line emission from UNSWIRF Field 1 of Figure 1,
showing the northern portion of the CND\@.  Contours are overlaid on a
greyscale representation of the same data. Contour levels start at,
and are in increments of, $5 \times 10^{-16} \u erg \ut s -1 \ut cm -2 
\ut arcsec -2 $.  (b) As for (a) except that the contours are overlaid on
a 2.2\micron\ continuum image of the region. Labelled, around the
edges of the image, are the line center velocities  in \kms\ (V$_{LSR}$) of
prominent features.  Also indicated with plus signs are the location
of OH 1720\,MHz masers, labelled with their associated emission
velocity, and with a corss the location of Sgr A$^*$.  The brightest
2\micron\ source, IRS\,7, lies 5'' to the N of Sgr A$^*$.
The central velocities of
emission features are labelled to the nearest 5 \kms.  
The plus signs show the position of OH masers.
}
\end{figure}

\begin{figure}
\caption{As for Figure 5, for  UNSWIRF Field 2 of Figure 1,
the southern portion of the CND\@. } 
\end{figure}

\begin{figure}
\caption{A contour of 20\,cm continuum emission 
outlining the Sgr A East shell,  shown  at 0.75 mJy/pixel,  
pixel scale 0.5$''$ and  resolution  3.1$''\times1.6''$, is
superimposed on the grayscale continuum-subtracted NICMOS image of the
2.12\micron\ H$_2$ 1--0 S(1) line emission from the CND and Sgr A East. 
The plus  sign coincides with the position of maser source C 
from Yusef-Zadeh \etal
(1996) and the cross sign coincides with the position of Sgr A$^*$.  }

\end{figure}

\begin{figure}
\caption{As for Figure 5b, for  UNSWIRF Field 3 of Figure 1,
to the SE of the CND\@. }
\end{figure}

\begin{figure}
\caption{As for Figure 5b, for  UNSWIRF Field 4 of Figure 1,
to the E of the CND\@. }
\end{figure}

\begin{figure}
\caption{A greyscale distribution of the ionized gas
emitting in the continuum at 1.2\,cm, obtained with the VLA at a resolution
of 0.3$''\times 0.2''$, showing the Northern arm of the Sgr A West
together with a shell of weak ionized gas around its northern tip.
In contract to all other grayscale images shown in 
this paper, this figure shows a positive grayscale 
as  the 
continuum emission is seen in white. The arrows are
drawn to
indicate the region of interest.}
\end{figure}

\end{document}